\documentclass[usenatbib,usegraphicx,usedcolumn]{mn2e}
\usepackage{bm}
\usepackage{amssymb,amsmath}
\usepackage{graphicx}
\usepackage{natbib}
\usepackage{program}
\usepackage{color}
\definecolor{darkgreen}{rgb}{0.15,0.5,0.15}
\definecolor{darkblue}{rgb}{0.15,0.15,0.5}

\newcommand\kpc[1]{~\rm{kpc}#1}

\usepackage[backref,breaklinks,colorlinks,citecolor=blue]{hyperref}

\title[Milky Way's halo and subhalos in SIDM]{The Milky Way's Halo and Subhalos in Self-Interacting Dark Matter}

\author[Victor H. Robles et al.]{Victor H. Robles\thanks{E-mail: roblessv@uci.edu}$^{1}$, Tyler Kelley$^{1}$, James S. Bullock$^{1}$,  Manoj Kaplinghat$^{1}$\\
$1$ Department of Physics and Astronomy, University of
       California, Irvine, 4129 Frederick Reines Hall, Irvine, CA 92697,
       USA\\}

\begin{document}
\date{MNRAS, xxx 2018}

\pagerange{\pageref{firstpage}--\pageref{lastpage}} \pubyear{2018}

\maketitle

\label{firstpage}

\begin{abstract}
We perform high-resolution simulations of a MW-like galaxy in a self-interacting cold dark matter model with elastic cross section over mass of $1~\rm cm^2/g$ (SIDM) and compare to a model without self-interactions (CDM). We run our simulations with and without a time-dependent embedded potential to capture effects of the baryonic disk and bulge contributions.  
The CDM and SIDM simulations with the embedded baryonic potential exhibit remarkably similar host halo profiles, subhalo abundances and radial distributions within the virial radius. The SIDM host halo is denser in the center than the CDM host and has no discernible core, in sharp contrast to the case without the baryonic potential (core size $\sim 7 \, \rm kpc$). The most massive subhalos (with $V_{\mathrm{peak}}> 20 \, \rm km/s$) in our SIDM simulations, expected to host the classical satellite galaxies, have density profiles that are less dense than their CDM analogs at radii less than 500 pc but the deviation diminishes for less massive subhalos. 
With the baryonic potential included in the CDM and SIDM simulations, the most massive subhalos do not display the too-big-to-fail problem. However, the least dense among the massive subhalos in both these simulations tend to have the smallest pericenter values, a trend that is not apparent among the bright MW satellite galaxies.
\end{abstract}
\begin{keywords}
cosmology: dark matter -- galaxies: halos -- MW
\end{keywords}

\section{Introduction}\label{sec:intro}

Self-interacting dark matter is a well-motivated idea that arises generically in hidden sector dark matter models~\citep{spe00,manoj16,tulin17}. It naturally retains all the successes of the cold dark matter model on large scales (\citet{springel16,trujillo11,vogel14a}), while modifying the dark matter halo density profiles within 5 to 10\% of the virial radius \citep{dave2001,vogel12,rocha13,peter13,zavala13,oliver15,vogel14b,fry15,robertson17,robles17}. 

These modifications to the halo density profile are in agreement with the diverse shapes of galactic rotation curves \citep{ren18,oman15,kamada17,creasey17} for cross section over mass, $\sigma/m$, in the range of $1-10 \, \rm cm^2/g$. They also alleviate the too-big-to-fail problem in the Milky Way, Andromeda and the Local Group \citep{boylan11,boylan12,tollerud14,kirby14,kimmel14,papa15,papa16}.

\begin{figure*}
\centering
\includegraphics[scale=.19,keepaspectratio=true]{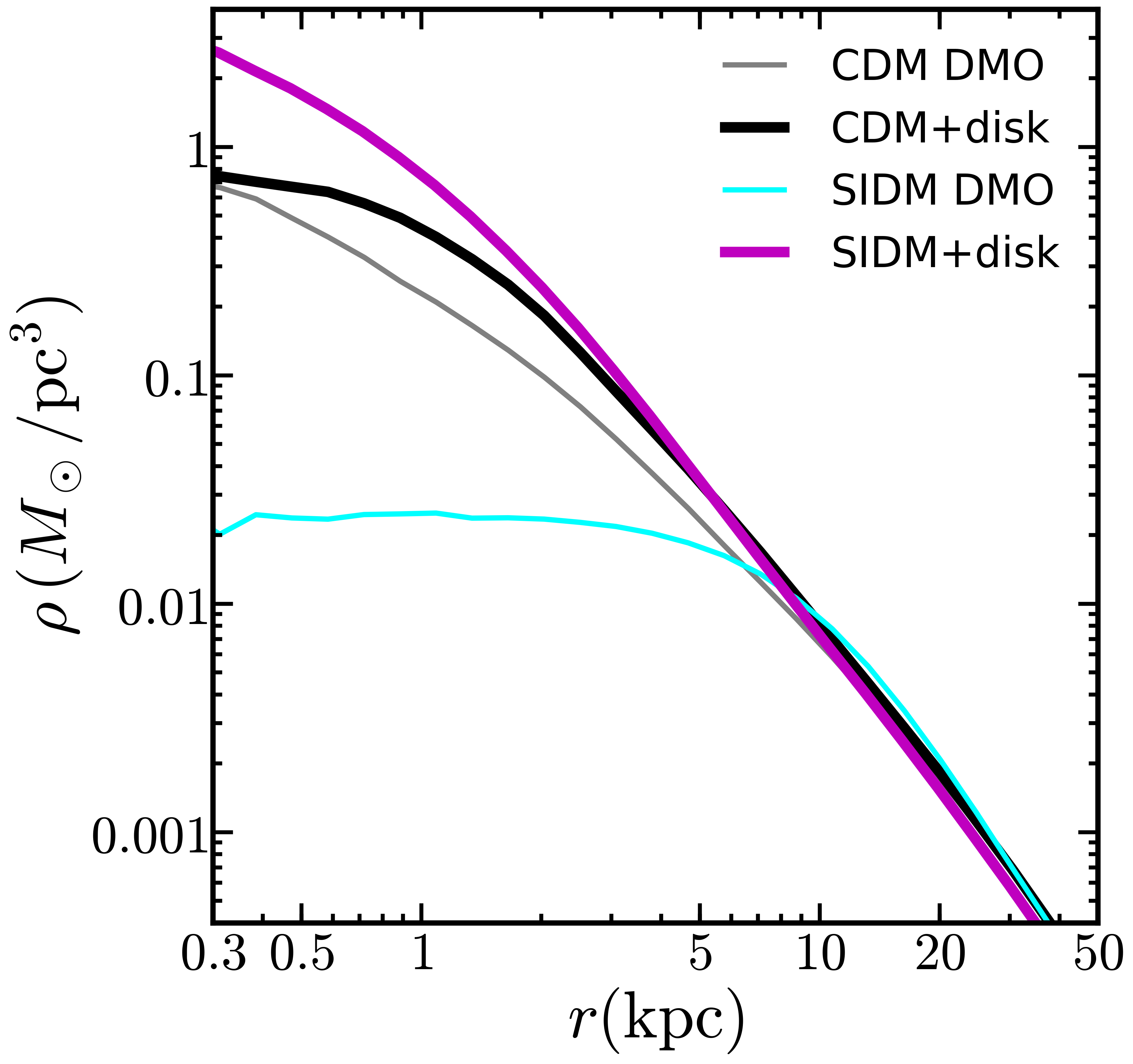} 
\includegraphics[scale=.19,keepaspectratio=true]{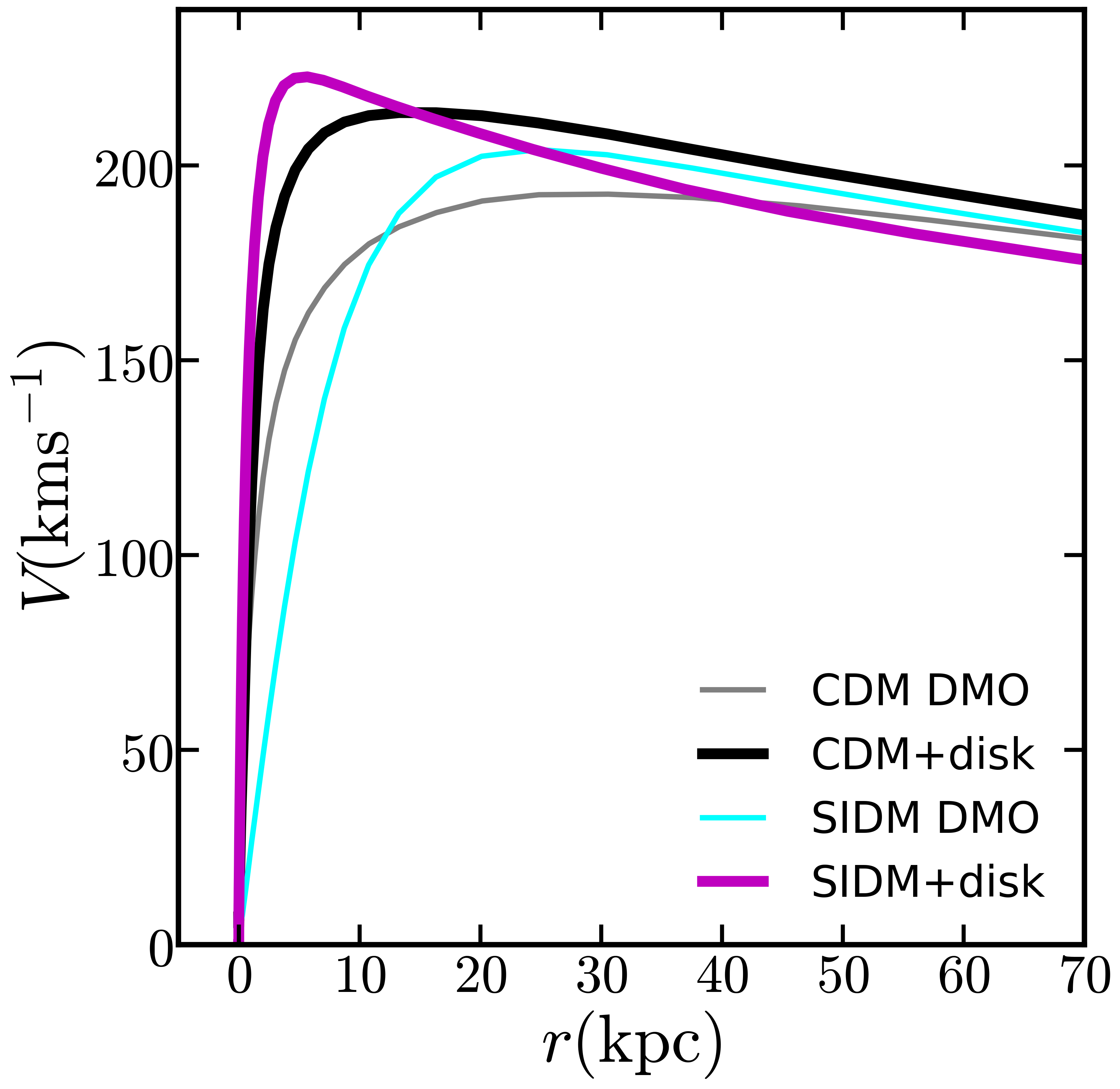} 
\includegraphics[scale=.19,keepaspectratio=true]{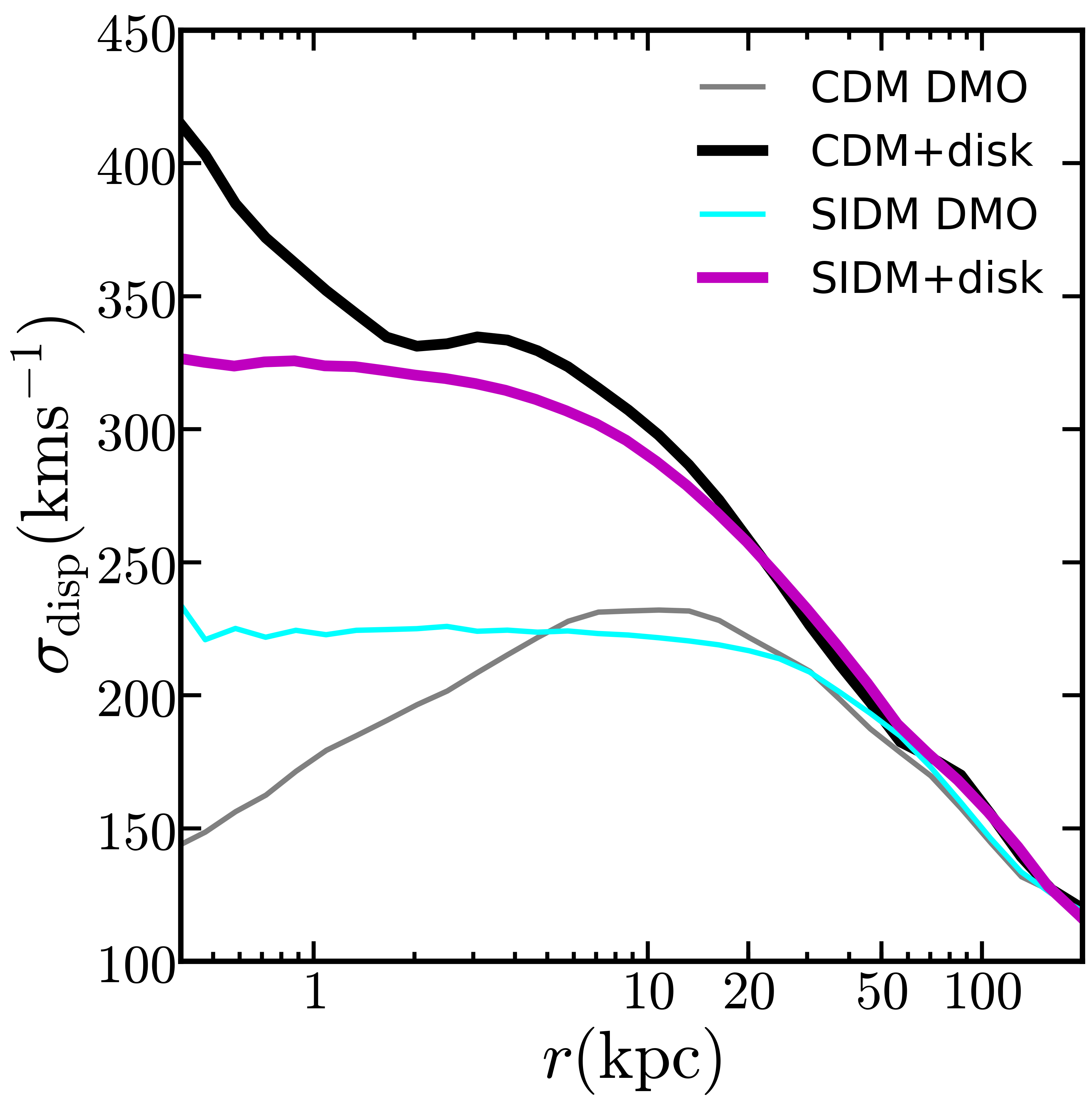} 

 \caption{
 Dark matter density (left), circular velocity (middle) and velocity dispersion profiles (right) for the MW host. In the SIDM DMO simulation (cyan) a large core develops due to the self-interactions thermalizing the center of the halo, whereas the collision-less CDM DMO (gray) simulation displays the usual central cuspy profile.  When the simulations include the contribution of the time-dependent baryonic potential the SIDM density profile (magenta) becomes denser than CDM (black). 
The stellar contribution has a strong impact on the DM velocity dispersion within the region of the disk in both CDM and SIDM simulations. The CDM+disk run shows an additional steep rise within $R_\star$ ($2.5 \, \rm kpc$ at $z=0$) in the velocity dispersion profile, which is not present in the SIDM+disk run because of the thermalization process.
}
\end{figure*}

Previous works have explored MW host properties and its response to baryons using a static potential for the baryons  \citep{oliver18,omid18}. \citet{dicintio17} performed SIDM hydrodynamical simulations to study the impact of super massive black holes, stars and gas on a MW-mass halo at redshift $z=0.5$, for $\sigma/m=10 \, \rm cm^2/g$.
Their DM simulation particle mass was $m_{\rm DM}=10^5M_\odot$, but a study of structural changes in low-mass halos requires even higher resolution, both in CDM \citep{fitts17} and SIDM \citep{robles17}.

In this letter we simulate CDM and SIDM MW-mass halos with an embedded time-dependent galaxy potential. We discuss the structural changes in the host and subhalo density profiles, and the implications for subhalo mass function and radial distribution. The SIDM model we simulate has $\sigma/m = 1 \, \rm cm^2/g$; our predictions should not be extrapolated to larger cross sections where the core sizes could be different (larger and smaller) because of the possibility of core collapse and the non-linear coupling between thermalization and tidal effects. 

The use of the baryonic potential overcomes the high computational cost that is required to resolve the structure of subhalos likely to host dwarfs with stellar masses of $M_\star \sim10^5M_\odot$ \citep{zolotov12,sawala16,wetzel16,kimmel18}.
\cite{kimmel17} show that embedding a time-dependent galaxy potential at the center of a high resolution MW-mass DMO simulation yields substructure populations in good agreement with fully hydrodynamical simulations of comparable resolution, but at  substantially less CPU cost.

We describe our simulations in section 2, section 3 describes the host dark matter density profile, sections 4 and 5 discuss the radial distribution and density profiles of the subhalos, in section 6 we explain the key features of the density profiles using an analytic model, in section 7 we discuss the orbits of the satellites and we conclude in section 8.

\begin{figure*}
\centering
\includegraphics[scale=.19,keepaspectratio=true]{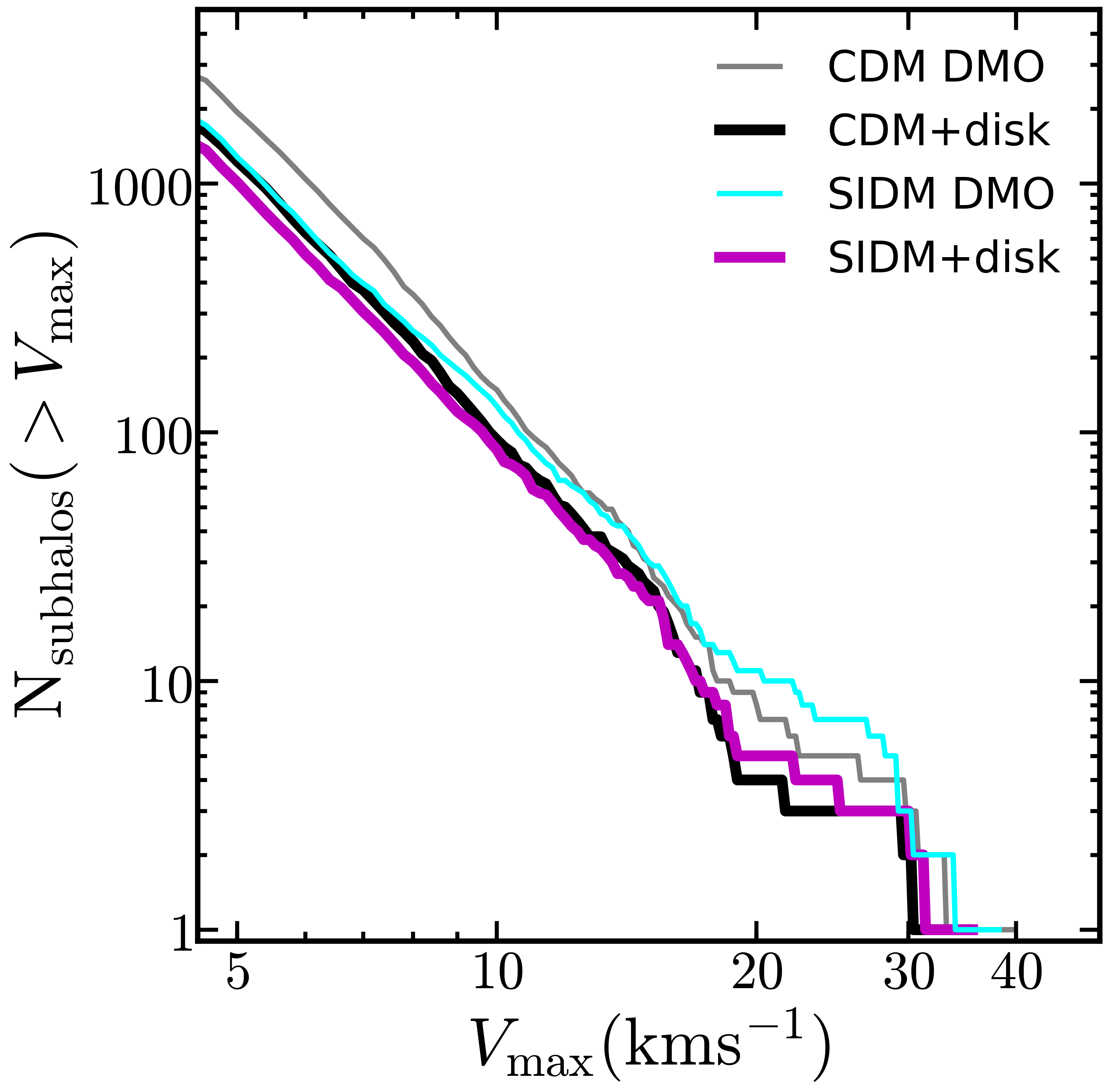} 
\includegraphics[scale=.19,keepaspectratio=true]{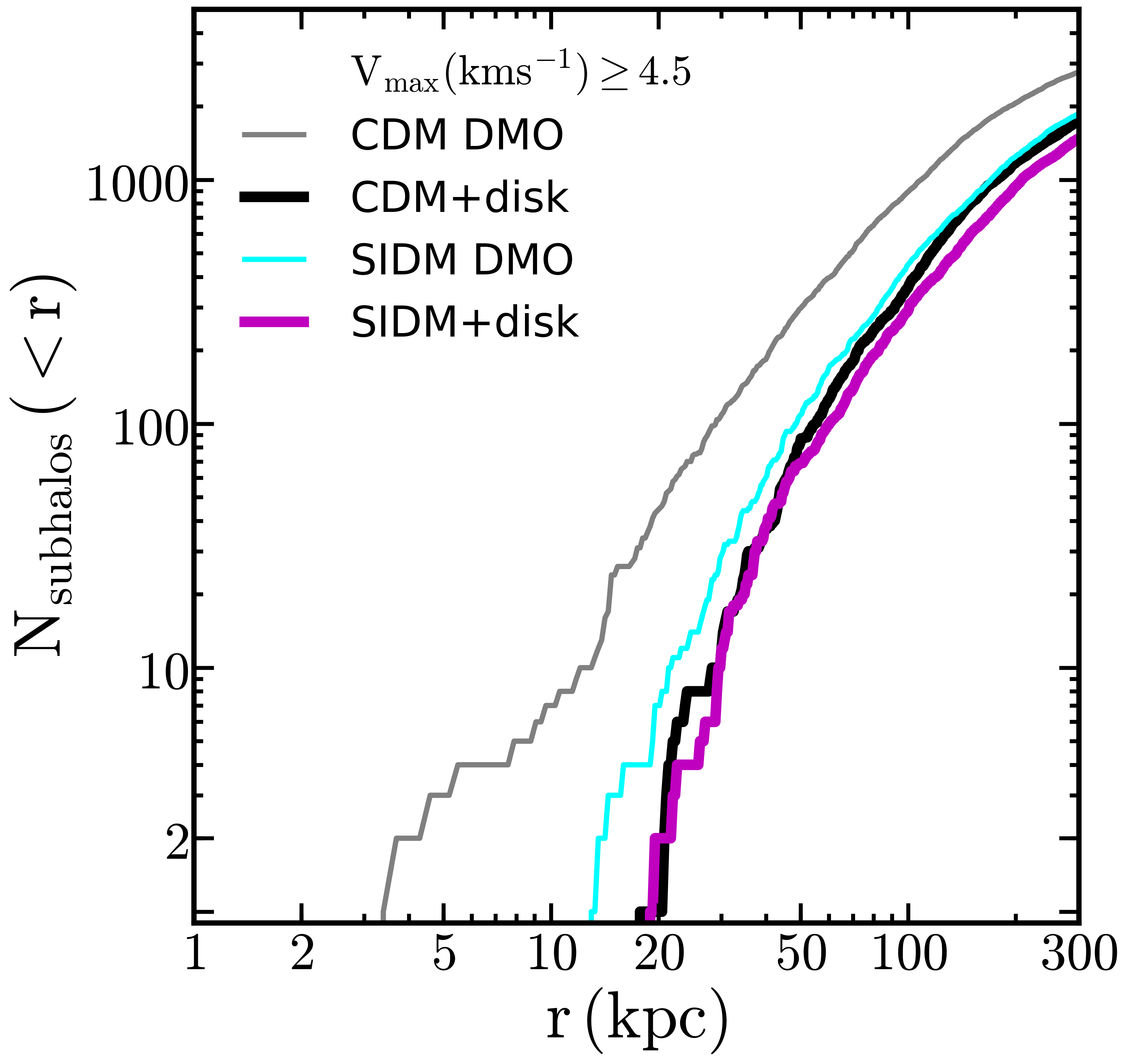} 
\includegraphics[scale=.19,keepaspectratio=true]{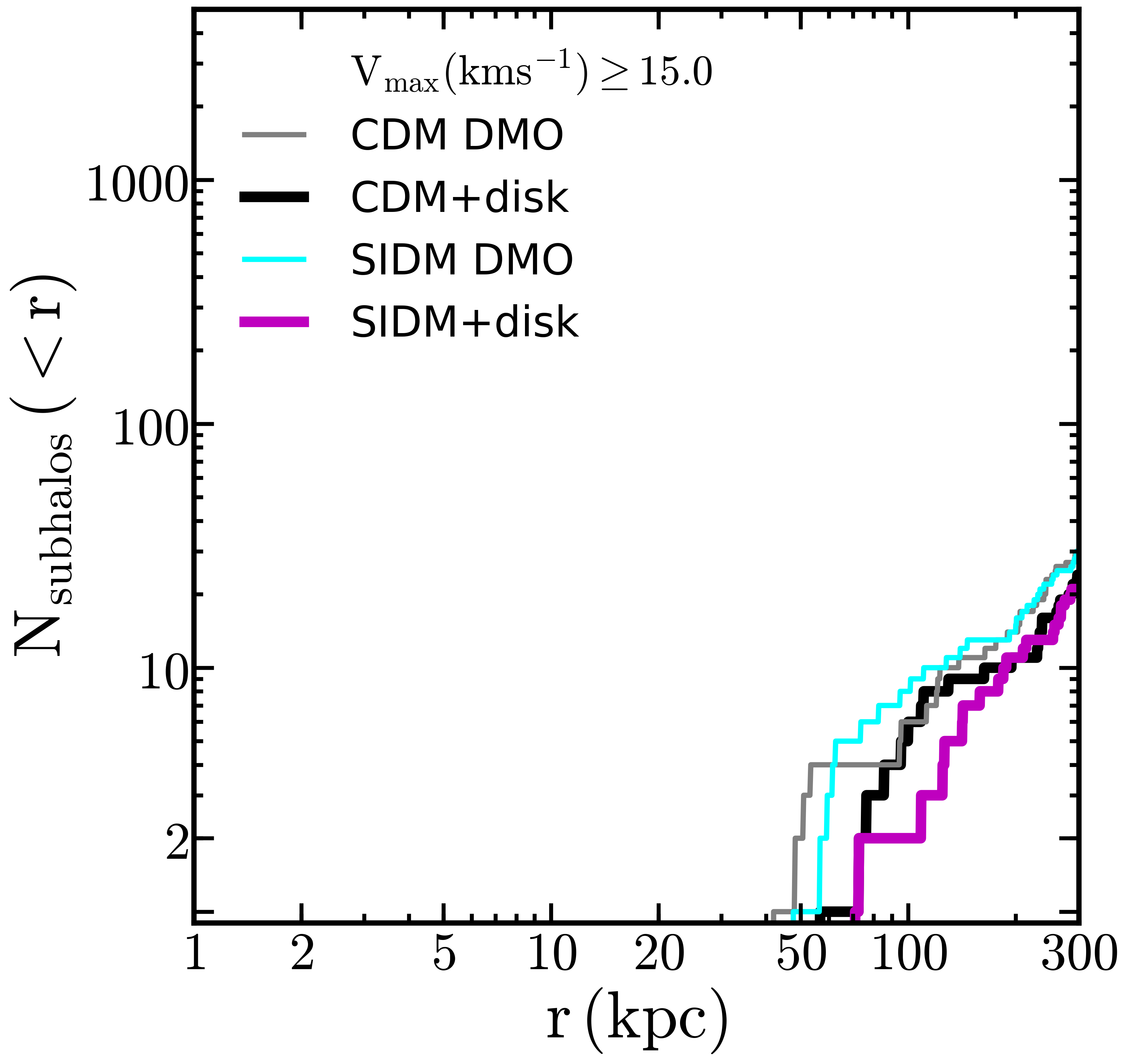}

 \caption{
 Left: Cumulative number of subhalos at $z=0$ above a given maximum circular velocity within $300 \kpc$ of the SIDM DMO (cyan), SIDM disk (magenta), CDM DMO (gray) and CDM disk (black) hosts. Middle: Cumulative counts of subhalos within a given radius for all resolved subhalos with $V_{\rm max}>4.5 \rm km s^{-1}$ (bound mass $ \sim 5 \times 10^6 M_\odot$). Right: Same as middle panel but for $V_{\rm max}>15 \rm km s^{-1}$. At $z=0$ and within $20 \kpc$ from the respective hosts, we find 2 surviving subhalos with $V_{\rm max}>4.5 \rm km s^{-1}$ in both CDM and SIDM disk simulations, and less than 10 subhalos in the SIDM DMO run, whereas there are $\sim 40$ in the CDM DMO simulation. 
Comparing the subhalo populations and radial distributions for the SIDM and CDM embedded disk runs does not reveal a substantial difference for $V_{\rm max}>4.5 \rm km s^{-1}$, or $V_{\rm max} \geq 15 \rm km s^{-1}$ subhalos.
}
\end{figure*}

\section{Simulations}\label{sec:sims}
We simulate 4 cosmological MW-mass halos: CDM DMO, CDM+disk\footnote{We will refer to the simulations that have an embedded baryonic potential as `disk' simulations for simplicity, however, these simulations also include a bulge component.}, SIDM DMO and SIDM+disk, all of them evolve in a Planck cosmology \citep{planck15}, with $h=0.675$, $\Omega_m=0.3121$, $\Omega_b= 0.0488$ and $\Omega_\Lambda=0.6879$. We apply the `zoom-in' technique \citep{katz93,onorbe14} to simulate a high-resolution region (with dark matter particle mass $m_{\rm DM} \approx 30,000 M_\odot$ and physical Plummer equivalent softening $\epsilon_{\rm DM}=37 \rm pc$) of the same cosmological volume of a box length of 50 $\rm Mpc/h$ such that the region contains a single MW-mass halo ($10^{12} M_\odot$) within $10 R_{\rm vir} \sim 3 \rm Mpc$ from the selected halo at z=0.

We run all simulations using the code \texttt{GIZMO} \citep{hopkins15}\footnote{http://www.tapir.caltech.edu/~phopkins/Site/GIZMO.html}, we identify halos using \texttt{Rockstar} \citep{behroozi13a} and use the halo catalogs that we build from the merger trees using \texttt{consistent-trees} \citep{behroozi13b}, all initial conditions were generated at $z = 125$ using \texttt{MUSIC} \citep{hahn11} with second-order Lagrangian perturbation theory. In the CDM+disk and SIDM+disk simulations, we include a baryonic disk following the same procedure described in \citet{kelley2018}  to generate initial conditions starting at $z = 3$ from the DMO simulations.   As in \citet{kelley2018}, our disks are composed of two components, gas and stellar, we additionally include a bulge component modeled by a Hernquist sphere \citep{hernquist90} with constant bulge-to-disk mass ratio set by present values, all the baryonic components are chosen to agree with $z=0$ MW observations. 

The stellar disk mass and disk size increase with time in the CDM+disk and SIDM+disk runs, the disk mass begins to grow at $z=3$ at the same rate as the halo mass following the abundance matching relation \citep{behroozi13}. 

The MW baryonic gas and stellar disks are modeled as exponential disks. In the simulations we achieve this, for each component, by using three overlapping \citet{miyamoto75} potentials that yield an effective exponential disk \citep{smith15}. The exponential radial and vertical scale lengths and total masses at z=0 of the stellar and gas disks are $R_\star=2.5 \kpc$, $h_\star=0.35\kpc$, $M_\star=4.1 \times 10^{10}M_\odot$ and $R_{\rm gas}=7 \kpc$, $h_{\rm gas}=0.084\kpc$, $M_{\rm gas}=1.86 \times 10^{10}M_\odot$, in agreement with current constraints \citep{hawthorn16}.
The bulge mass and scale radius at $z=0$ are $9 \times 10^{9} M_\odot$ and $0.5 \kpc$. 

The dark matter self-interaction implementation is the same as the one used in \citet{rocha13,robles17}, i.e., we assume identical dark-matter particles undergoing isotropic, velocity-independent, elastic, hard-sphere scattering with a cross section $\sigma$. Our SIDM runs assume $\sigma/m=1 \, \rm cm^2/g$ to maximize any difference with CDM while remaining in the range of current constraints \citep{manoj16}.
\begin{figure*}
\centering
\hspace{-1cm}
\includegraphics[width=6.4cm, height=6.4cm]{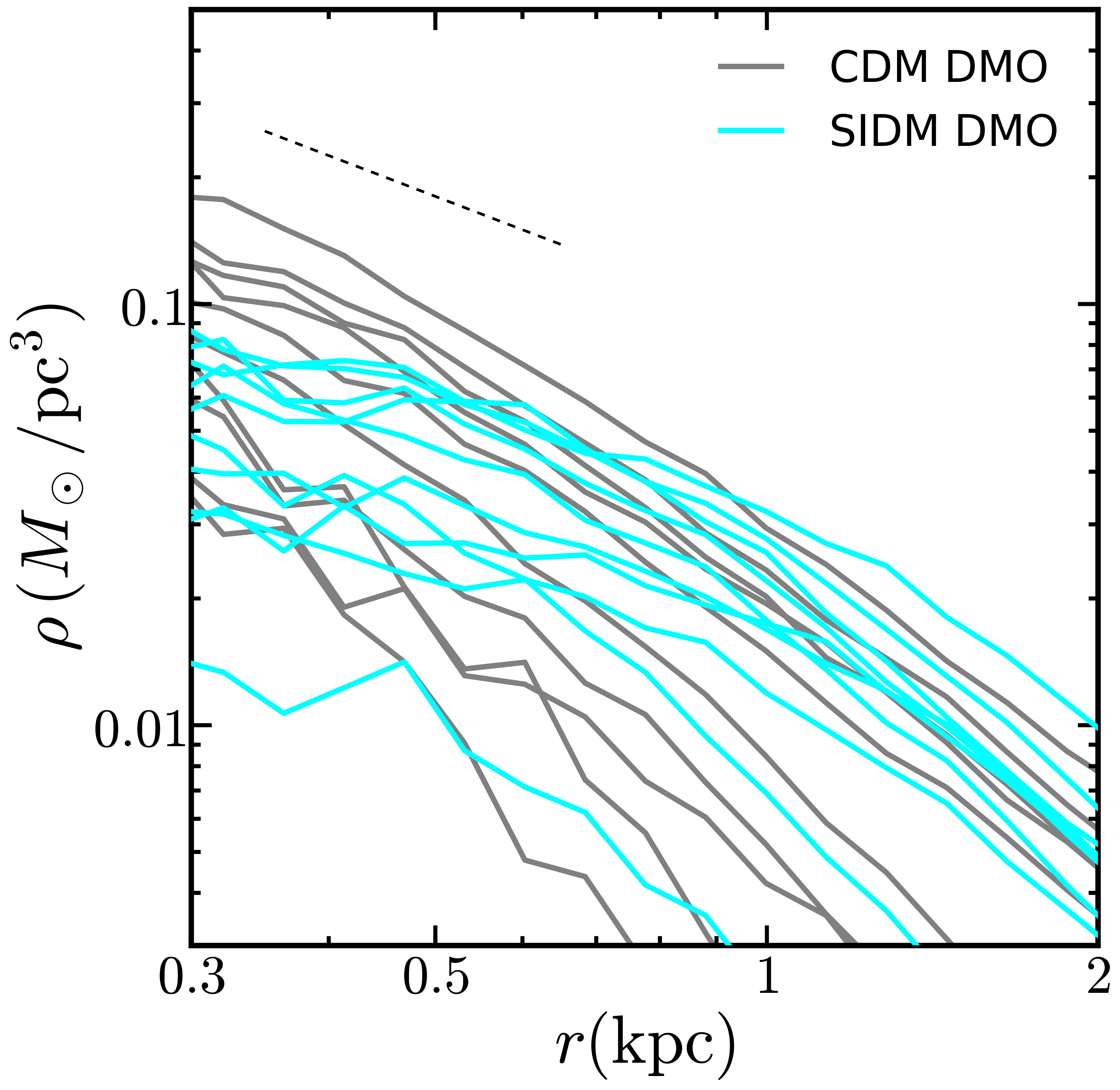}
\hspace{.2cm}
\includegraphics[width=6.4cm, height=6.4cm]{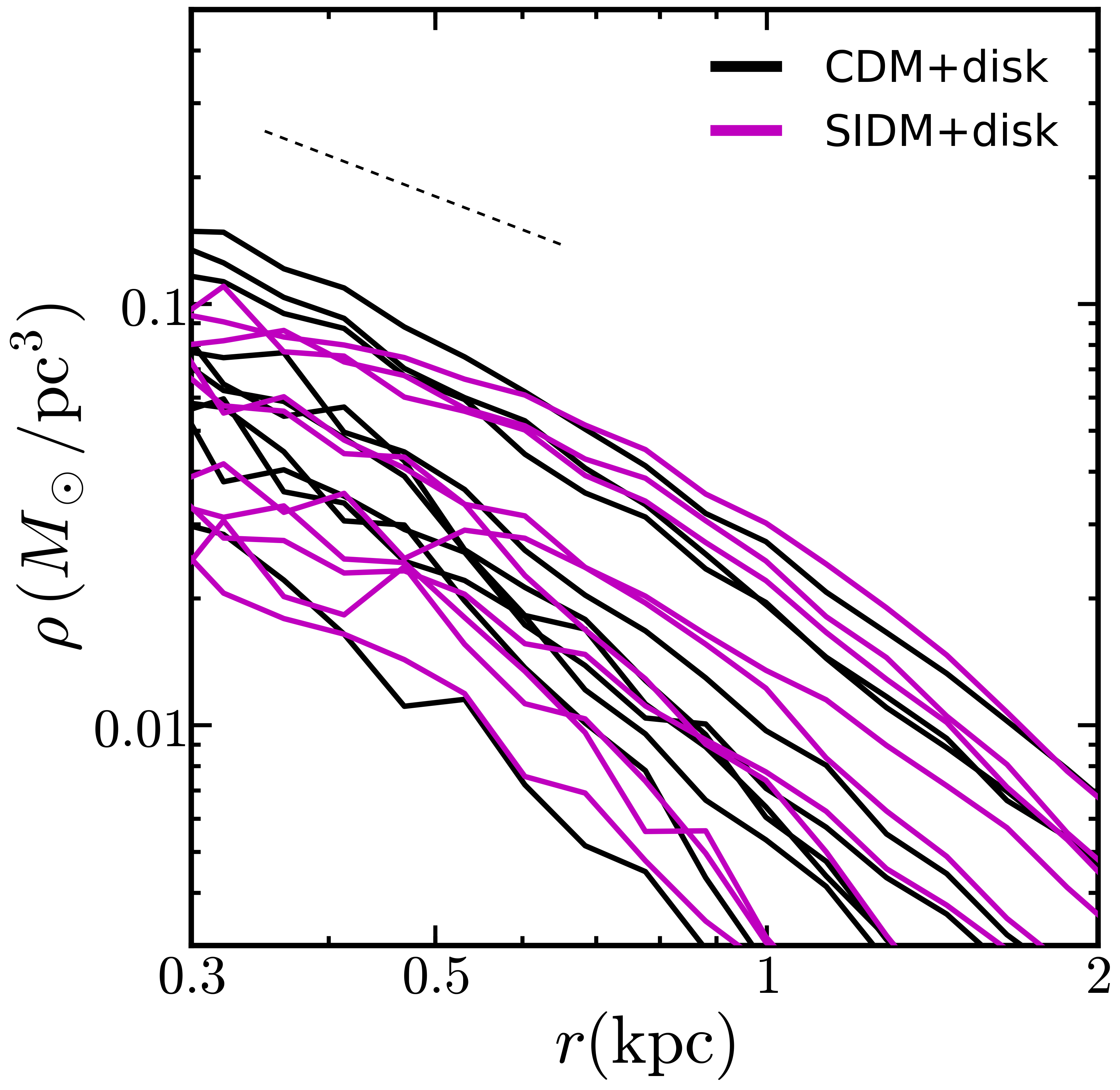} \\
\includegraphics[width=6.8cm, height=6.5cm]{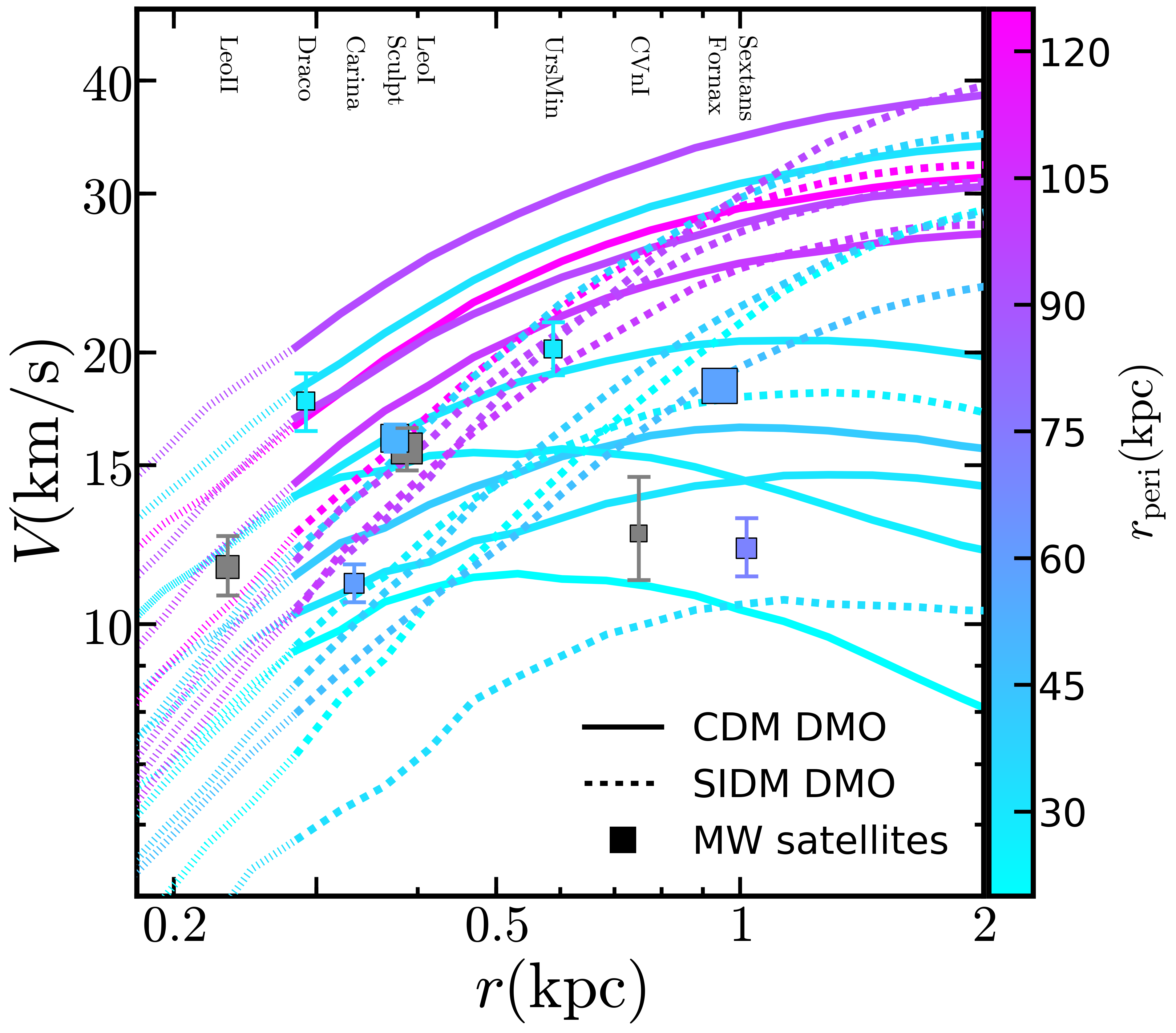}
\hspace{.2cm}
\includegraphics[width=6.8cm, height=6.5cm]{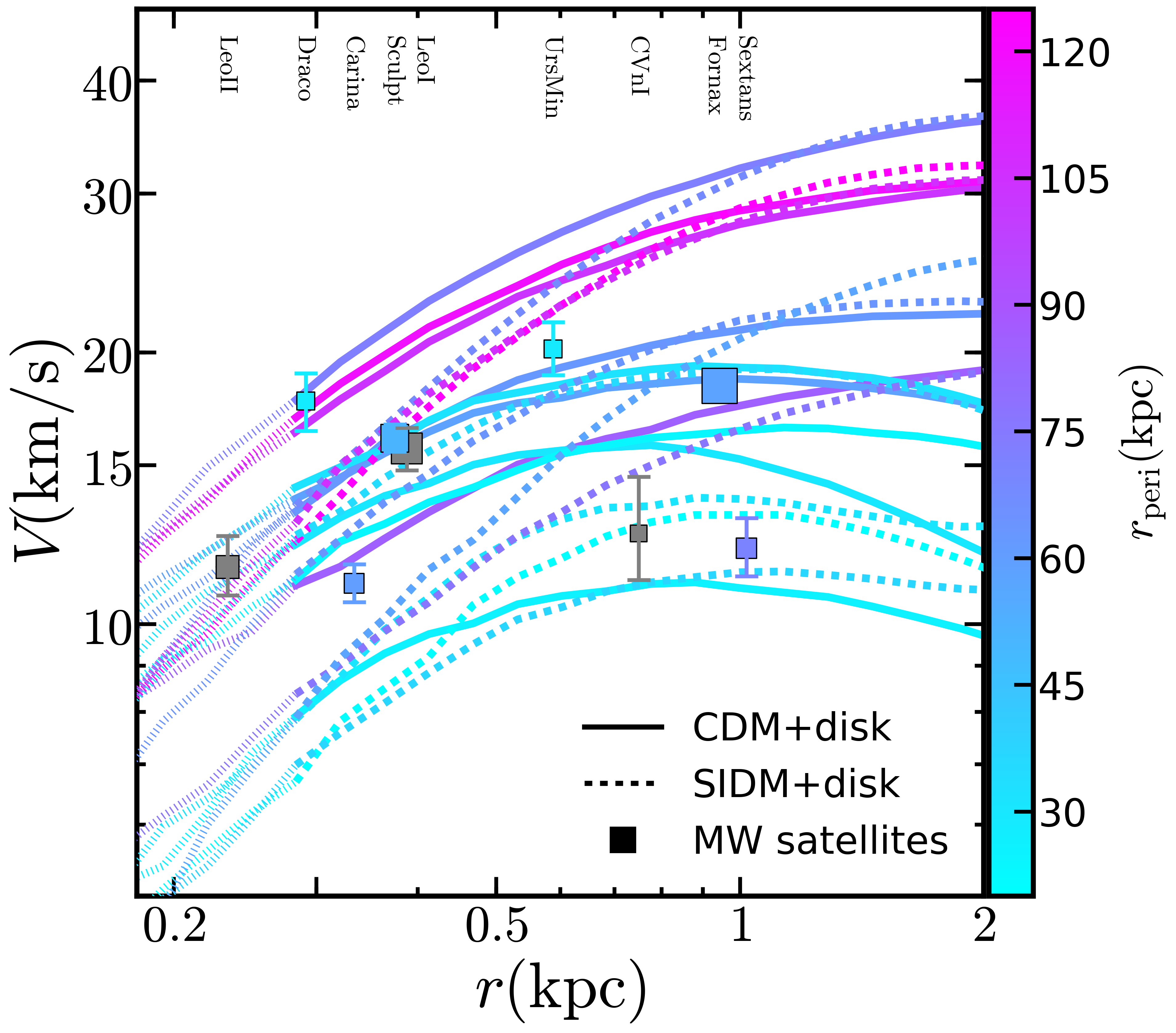} 
 \caption{
 Comparison of the density (upper row) and circular velocity (bottom row) profiles for the ten subhalos with the largest peak velocity in the simulations with the disk and bulge potentials (right) and the DMO simulations (left). The five densest subhalos in the SIDM DMO simulations have core sizes $\sim 400$ pc and are less dense than the five densest subhalos in CDM, whose central densities increase as $\rho \propto r^{-1}$ (as shown by the gray dashed line). A similar, but milder, difference in the central densities is also observed in the simulations including the disk and bulge potentials. While in the DMO runs SIDM has less scatter in the density profiles than CDM, this is increased in the presence of the embedded potential (top panels). In the bottom panels, squares are MW satellites brighter than $2 \times 10^5  \, \mathrm{L_\odot}$ \citep{wolf10}; the point sizes are proportional to the logarithm of their stellar masses and are color coded by the inferred pericenters given in \citep{fritz18} using \texttt{Gaia} DR2. Gray symbols are satellites whose inferred pericenters we consider unreliable, having errors of the same order as their mean value. The circular velocity curves for the simulated CDM (solid) and SIDM (dashed) subhalos are color coded by their pericenters, and the dotted lines indicate unresolved regions in our simulations. 
While our CDM halo halo has three subhalos as dense as Draco, the SIDM with $\sigma/m=1 \, \rm cm^2/g$ does not have any.
 In both dark matter models, subhalos with smaller pericenters (cyan lines) have lower densities. When the embedded potential is present, the same trend holds. However, the data do not seem to follow this trend.}

\end{figure*}

\section{Density profile of the Milky Way halo}\label{sec:discs}

In the absence of baryons, self-interactions (SI) form a $\sim 7 \kpc$ core in the Milky-Way mass halo (Figure 1), and its inner halo density is an order of magnitude smaller than the CDM DMO host density at $1$ kpc. However, we find that SIDM has no inner core and it is actually $\sim 2-3$ times denser within $\sim 1$ kpc than CDM (and a factor of $\sim 5$ denser than CDM DMO) when the baryonic potential is included. This was predicted by \citet{manoj14} and is a consequence of thermalization in the presence of the baryonic gravitational potential. 

The response to the baryonic potential is shown in Fig. 1. The contraction leads to a higher rotation velocity maximum in SIDM, but the changes are too small to a detectable difference given the uncertainties in the concentration of the halo, modeling the baryons and inferring the rotation curve in the inner parts. 
The velocity dispersion, $\sigma_{\rm disp}$, shows a larger variation between the two DM models. For SIDM+disk, $\sigma_{\rm disp}$ rises mildly within $R_\star$, and this negative temperature gradient leads to outward heat flow and an increasing central density (gravothermal collapse), which has been noted before \citep{oliver18}. 
Notably, we find CDM has a stronger response to the stellar contribution than SIDM; we find that the cold center observed without baryons (CDM DMO) has been dynamically heated by the presence of the stellar potential and there is a steep upturn in its velocity dispersion at $\sim R_\star$. It would be interesting to compare this expectation to a full hydrodynamical CDM simulation.

\section{Mass function and radial distribution of subhalos}
In Figure 2, we show the abundance of subhalos within $300~\rm kpc$ (about the virial radius) of the host, as a function of the maximum subhalo circular velocity, $V_{\rm max}$ (left panel). 
We observe only small differences in the total number of subhalos between the SIDM and CDM disk runs once the galaxy potential is included. 
A larger difference occurs in the DMO runs, where SIDM has less substructure than CDM for smaller $V_{\rm max}<15 \, \rm kms^{-1}$ subhalos. 

Comparing the number of massive subhalos ($V_{\rm max} > 15 \, \rm kms^{-1}$, which would host the bright MW satellites) in SIDM and CDM for simulations that include the galaxy potential, we find remarkably similar abundances. 
The small difference in the subhalo populations above $V_{\rm max}>15 \, \rm km s^{-1}$ is consistent with variations in the individual subhalo orbits arising from the non-linear evolution; we find that none of these subhalos had pericenter distances $r_{\rm peri}<15 \, \kpc$ ($r_{\rm peri}<20 \, \rm kpc$) for the DMO (disk) runs.

We also show the radial distribution for all well-resolved subhalos $V_{\rm max}>4.5 \rm  \, kms^{-1}$ (equivalent bound mass $ > 5 \times 10^6 \, M_\odot$) and more massive subhalos $V_{\rm max} > 15 \rm \, kms^{-1}$ in Fig. 2. The radial distributions of subhalos are very similar in CDM and SIDM once the baryonic potential are included.
This is consistent with the disk preferentially destroying subhalos that have orbits with small pericenters. 
In particular, the distribution of the most massive subhalos ($V_{\rm max} > 15 \rm \, kms^{-1}$) is remarkably consistent across the four simulations. For lower mass halos ($V_{\rm max} > 4.5 \rm \, kms^{-1}$), the presence of the baryonic potential mutes the differences between CDM and SIDM, indicating that number counts of subhalos alone will not be sufficient to constrain cross sections $\sigma/m < 1 \, \rm cm^2/g$.

\section{Density profiles of the satellites}
Our simulations resolve the inner $\sim 300$ pc of the most massive subhalos. 
Figure 3 shows the $z=0$ density profiles and circular velocity profiles of ten subhalos with the largest peak circular velocity ($> 20\, \rm km/s$) for the DMO and disk runs. 
In the CDM simulations, the subhalos are cuspy as expected and the influence of the disk does not substantially alter their inner densities. The data points in Fig. 3 are from inferences of the dynamical mass 
at the half-light radius $r_{1/2}$ for the MW dwarf spheroidals (dSphs) (taken from \citep{wolf10},  who  used  data  from  \citep{walker09,munoz05,simon07,mateo08,koch07}). 

The impact of the disk on the central densities of the dwarf spheroidals has been proposed as the solution to the too-big-to-fail problem, in conjuction with stellar feedback \citep{zolotov12,brooks14,sawala16}. Our work provides a clean demonstration of the impact of {\em just the disk} on the central densities of the subhalos, and it shows that the tidal effects of the disk can reduce the circular velocities at $300 \rm pc$ by 20-30\% for most subhalos, thereby alleviating the too-big-to-fail problem. 

In our SIDM DMO (disk) run, the five (three) most massive halos ($M>10^9 M_\odot$) develop a mild central core within $r<$ 400 pc, these cores have comparable sizes to those found in cosmological dwarf halos in the field for the same value of $\sigma/m$ \citep{robles17}. 
The lower mass SIDM subhalos have smaller cores and look cuspy over the range of scales that are resolved in our simulations. For $\sigma/m = 1 \, \rm cm^2/g$, most of the SIDM subhalos that host the classical dwarf spheroidals would have core sizes $< 300 \rm pc$ (unresolved in our simulations). MW dwarf spheroidals with stellar half-light radii $\sim 300$ pc \citep{mcon12} could be the best targets to discriminate between SIDM and CDM. In fact, the high density inferred for Draco cannot be explained by the 
subhalos in our $\sigma/m=1 \, \rm cm^2/g$ simulation (with or without the baryonic potential), consistent with the recent results of \citet{read2018}.

In addition to the dynamical mass estimates of the MW dwarfs, properties like stellar mass, size, star formation history, chemical enrichment and metallicity spread could be different in SIDM and CDM models, perhaps providing further handles on DM particle physics. 
This motivates running hydrodynamical SIDM simulations and their CDM pairs.

\section{Analytic SIDM density profile}
A simple working model for the SIDM halo profile is that the inner profile is isothermal, while the outer parts are the same as in CDM \citep{rocha13,manoj14}. Interpolating between the two profiles by smoothly joining the mass profile at a radius where (on average) SIDM particles have interacted once does a good job of matching simulation results for {\em field} halos that aren't in the {\em core collapse} phase \citep{manoj16,oliver18,omid18,robertson2018}. As expected the SIDM DMO density profile is almost perfectly reproduced by the \citet{manoj16} analytic model. We also found that the same model explains the core densities of subhalos reasonably well (see Appendix for details).  

In the SIDM+disk run, the host halo has developed a negative temperature gradient (core density is increasing) and the analytic model underpredicts the densities at intermediate radii, as expected~\citep{oliver18}.
Interestingly, matching the isothermal profile exactly to the NFW profile fails for these core-collapsed systems. 
Taking a cue from that, we solved the Jeans equation (starting with a constant density core at a small radius as before) allowing for the 1-d SIDM dispersion to decrease mildly with radius. We found the smallest gradient to get an exact match between the density and mass of the solution to the Jeans equation and a NFW profile. Using this gradient resulted in a good match to the simulated SIDM+disk profile (see Appendix for details). The generality of this modification  remains to be investigated.

\section{Orbits of satellites}
Figure 3 shows both the data and simulations colored by their pericenters of the satellite and subhalo orbits. We use the inferences of the pericenter values for the dSphs using \texttt{Gaia} data from \citet{fritz18} for their Milky Way halo mass of $8\times 10^{12}M_\odot$, which is close to our simulated halo. 

The comparison of the orbits of the subhalos to the measured pericenter distances from \texttt{Gaia} for the satellites presents an interesting puzzle. In the presence of the disk, the central density of the CDM subhalos is reduced (as expected). In this case, the two densest satellites Draco and Ursa Minor have to be hosted in halos that have pericenter distances at the upper end of the distribution. However, \texttt{Gaia} measurements suggest that these two satellites have likely come as close as 30 kpc to the center of the halo. On the other hand, the lower density satellites (e.g. Sextans) are hosted in subhalos that have been significantly affected by tidal interactions (smaller pericenters). Statistically assessing the too-big-to-fail problem in CDM and SIDM models and determining whether the mismatch between the observed satellite pericenters and the simulated subhalos is a strong discrepancy are issues that remain to be addressed.

\section{Conclusions}

We have presented results from high-resolution cosmological simulations of a Milky Way like galaxy in models with no dark matter self-interaction (CDM) and self-interaction cross section over mass $1\, \rm cm^2/g$ (SIDM). We included a realistic contribution of the stellar disk and bulge with an evolving embedded potential that grows in mass and size from $z=3$ to the present. The simulated disk and bulge properties at $z=0$ agree with MW observations.

We find that in the absence of a baryonic potential, a MW halo with dark matter self-interaction strength $\sigma/m=1 \, \rm cm^2/g$ develops a large core at its center and is nearly ten times less dense within $1\, \rm kpc$ than its corresponding CDM DMO halo. When the disk and bulge potential is included, the behavior is inverted: the SIDM halo is a factor of 2-3 denser than its equivalent CDM halo. This behavior can be understood by considering the thermalization of the inner region. 

We find the substructure abundance and the overall radial distribution of subhalos are similar once the baryonic potential is included suggesting that matching the number of observed satellites via abundance matching would yield same results for SIDM and CDM. 

We find that the internal structure of the most massive subhalos in both DMO and disk SIDM simulations reveal shallow profiles within about $500 \, \rm pc$, comparable to the effective stellar half-light radii of some MW and M31 dwarf spheroidal satellite galaxies \citep{mcon12,tollerud14}. The gravitational impact of the disk is sufficient to lower the simulated subhalo densities within a kpc to be consistent with the range of densities inferred for the MW dwarf spheroidals. The lowest  density subhalos are those with the smallest pericenters, a trend that doesn't seem to be reflected in the measurements of densities and pericenters of the bright MW satellites. 

None of the subhalos in the SIDM simulation was dense enough to match the inferred density of Draco. Our results suggest a detailed comparison of the stellar properties of classical dwarf spheroidals, including their orbital histories, could be used to discriminate between the two models. 
Including the ultra-faint satellites of the Milky Way, the dwarf spheroidals in Andromeda and the other Local Group dwarf galaxies should further sharpen this test.

\section*{Acknowledgments}

VHR was supported by a Gary A. McCue postdoctoral fellowship.  JSB and VHR were supported by NSF AST-1518291, HST-AR-14282, and HST-AR-13888. TK and JSB were  supported by NSF AST-1518291, HST-AR-14282, and HST-AR-13888. MK was supported by NSF Grant No. PHY-1620638. The simulations were run on the Texas Advanced Computing Center (TACC; http://www.tacc.utexas.edu), the NASA Advanced Supercomputing (NAS) Division and the NASA Center for Climate Simulation (NCCS), and the Extreme Science and Engineering Discovery Environment (XSEDE), which is supported by National Science Foundation grant number OCI-1053575.

\begin{figure*}
\centering
\includegraphics[scale=.25,keepaspectratio=true]{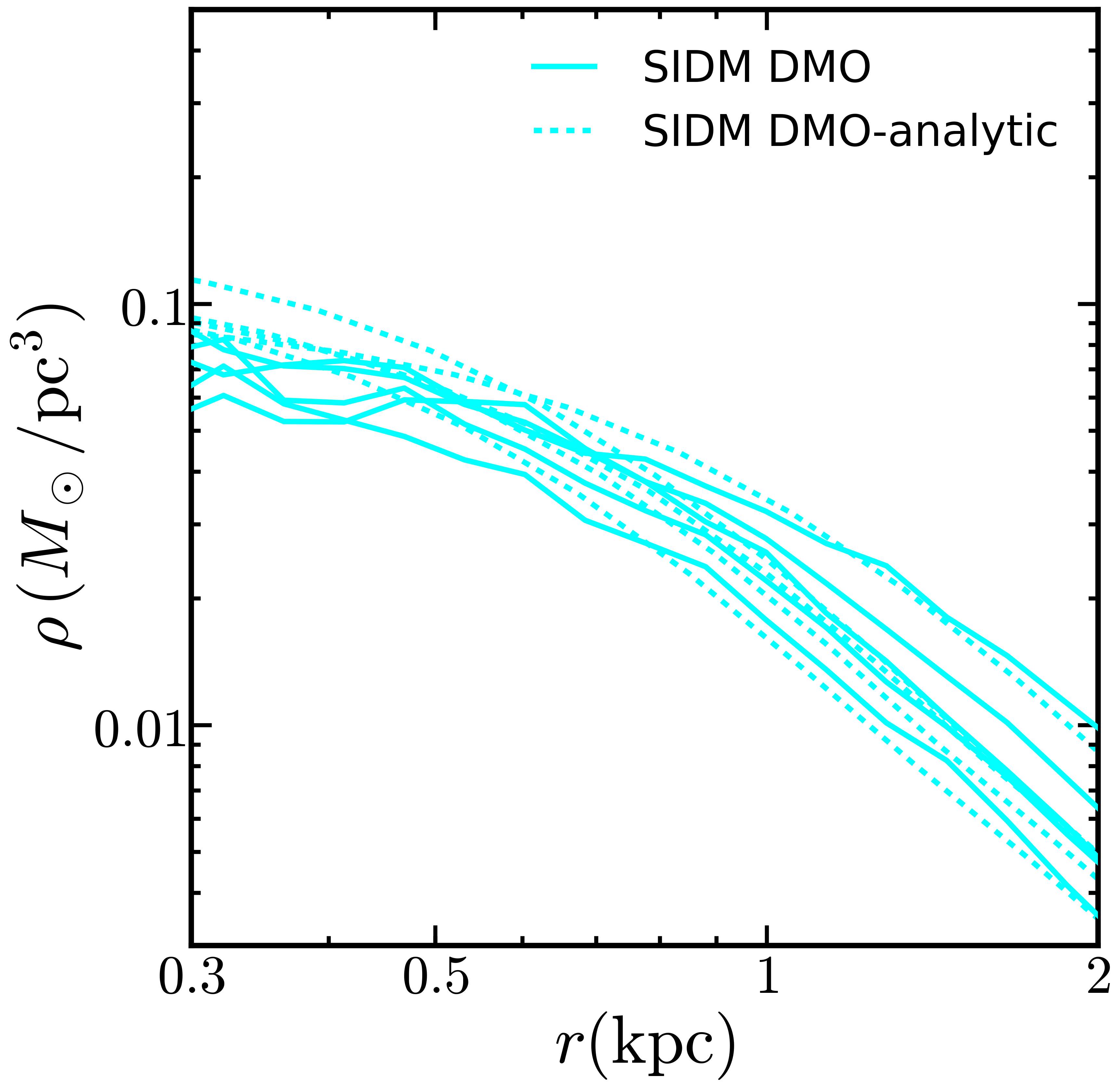} 
\includegraphics[scale=.25,keepaspectratio=true]{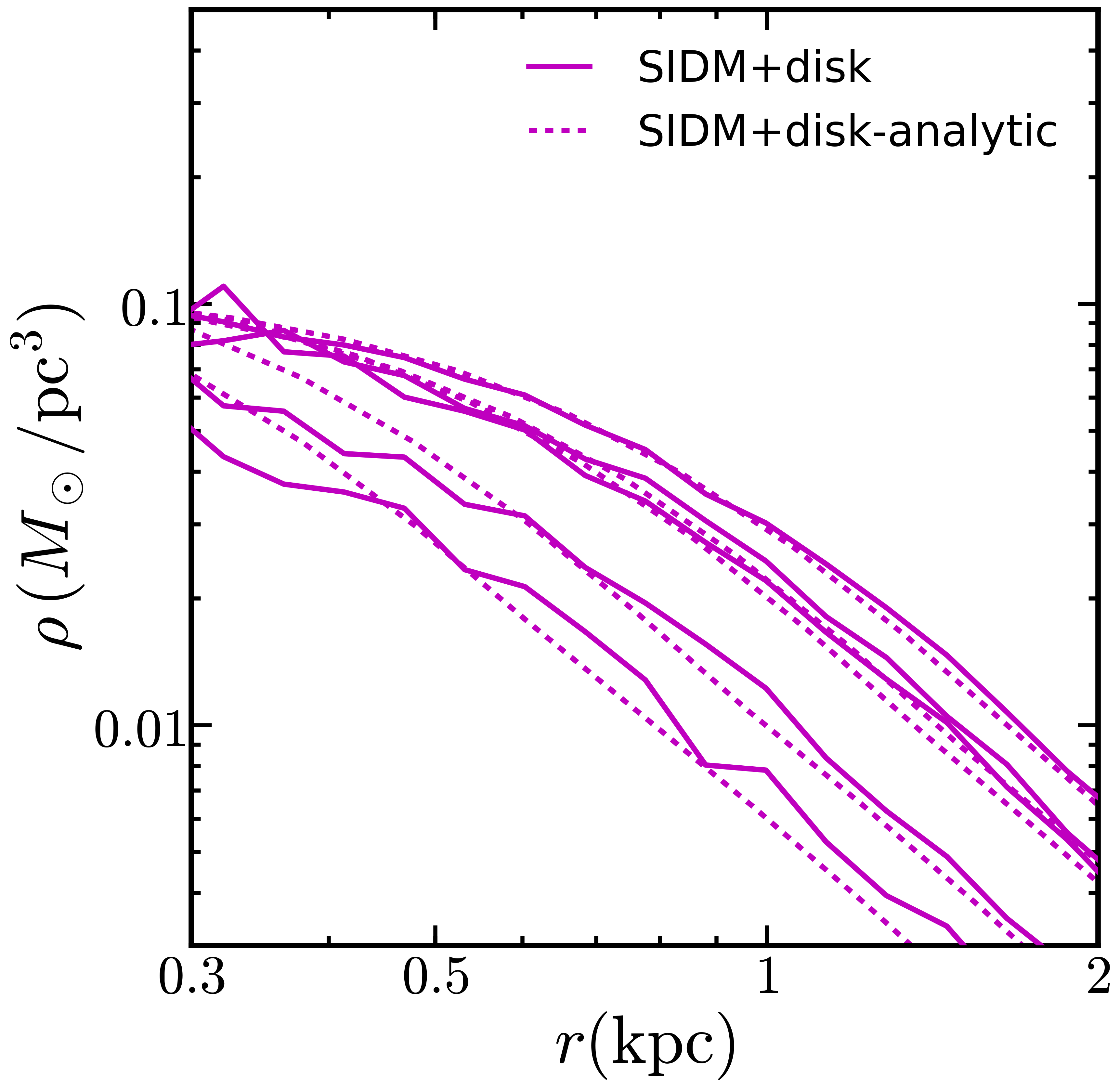} 

 \caption{Density profiles for the five densest subhalos at redshift $z=0$ (solid lines) along with their analytic fits (dotted lines) computed from the CDM subhalos following the procedure in \citet{manoj16}. The inferred analytic profiles from CDM subhalos are  less than $\sim 10\%$ different from the simulated SIDM subhalo profiles. This implies that core formation in SIDM subhalos can be understood as the consequence of DM thermalization caused by  self-interactions within a region where the average scattering rate per particle times the halo age is $\sim 1$.}
\end{figure*}

\begin{figure}
\centering
\includegraphics[scale=.25,keepaspectratio=true]{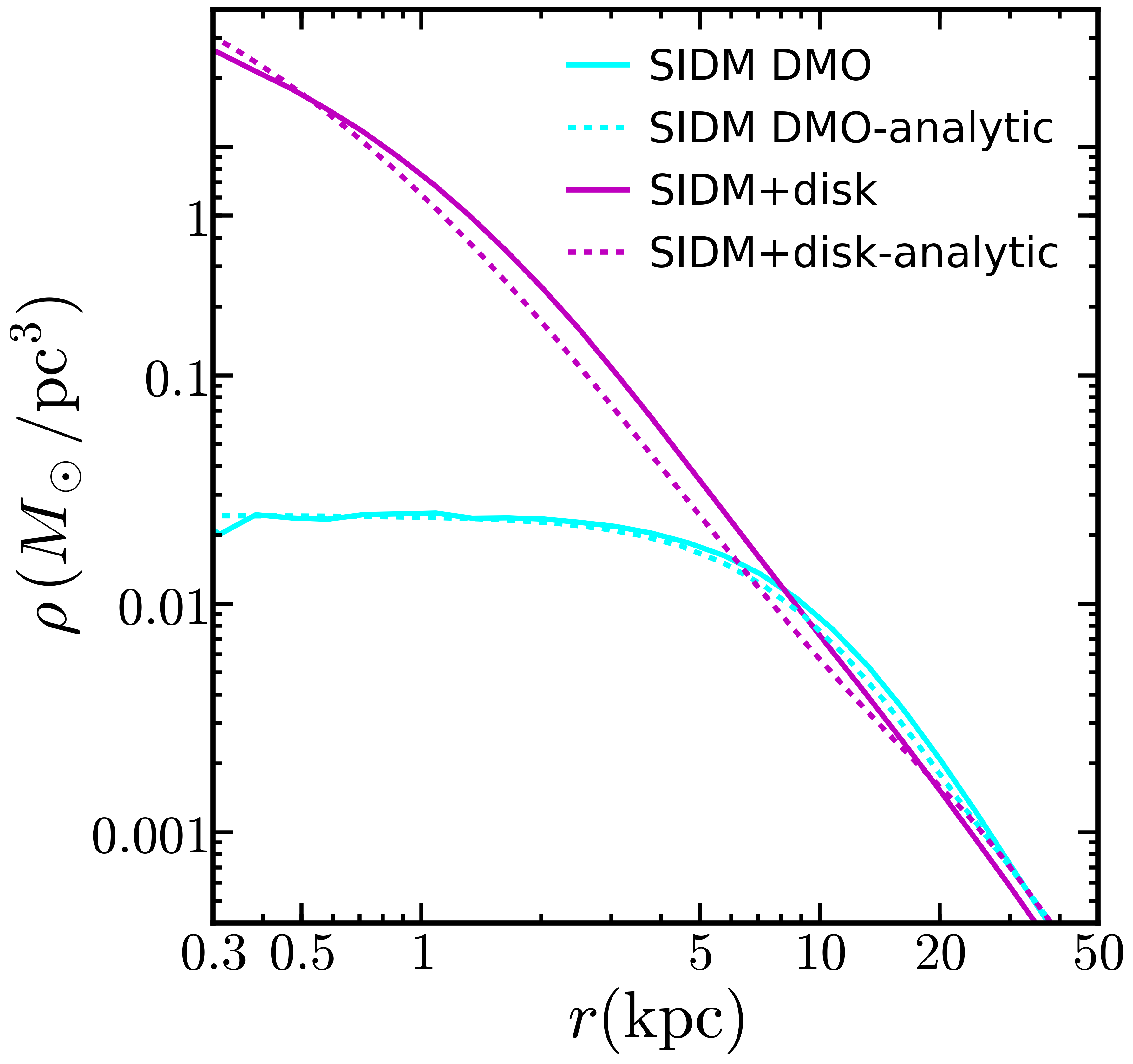} 
 \caption{Analytic (dotted lines) SIDM density profiles of MW host halo for the DMO (cyan) and embedded baryonic potential (magenta) cases. The SIDM MW halo profiles from the corresponding simulations is also shown (solid lines) for comparison. We observe good agreement between the profiles in the DMO simulations. For the SIDM+disk simulation, the fit is good in the baryon-dominated ($\leq 1\kpc$) and DM-dominated regions ($>10 \kpc$); the fit worsens in the transition region but the deviations are below about $25\%$.}
\end{figure}

\section*{Appendix: Analytic fits to SIDM halos}

The halo profile of SIDM halos can be understood using a simple analytic model\citep{manoj16}. The inner profile is isothermal, which implies that the halo density $\rho(r) \propto \exp(-3\Phi(\vec{r})/\sigma_{\rm disp}^2)$ where $\sigma_{\rm disp}$ is the 3D velocity dispersion. In the region where interactions have not had time to thermalize the halo, the density profile will be the same as in CDM. The CDM profile will be close to the NFW profile in regions sufficiently far from the influence of baryons. This picture works well for $\sigma/m \gtrsim 1 \, \rm cm^2/g$. 

A simple ansatz for interpolating between the isothermal and NFW profiles is to match the mass and density at a radius $r_1$ where the halo particles have had on average one interaction. Thus, $r_1$ is defined by the relation $\Gamma(r_1) \times t_{\rm age}=1$, where $t_{\rm age}$ is halo age. 
We show the results from this analytic model (assuming $\sigma/m=1 \, \rm cm^2/g$ and $t_{age}$=13 Gyr) in Fig. 4 (dotted lines) obtained by matching the isothermal solution to NFW fits to the CDM subhalos in Fig. 3. We used the {\tt Colossus} Python package to obtain the NFW fits. The analytic profiles are compared to the subhalo profiles in the simulations with and without the baryonic potential. The SIDM analytic profiles inferred from CDM subhalos are remarkably similar (less than $\sim 10\%$ difference) to the simulated SIDM profiles, showing that the core formation in SIDM subhalos can be understood from the thermalization of the dark matter caused by self-interactions.

It was pointed out by \citet{manoj14} that the isothermal requirement in the presence of a sufficiently compact stellar disk (or a massive bulge) would drive the SIDM density profile to be dense (thereby reducing the core size). The analytic model discussed above was shown to work well for N-body SIDM simulations with a static disk  \citep{oliver18,omid18}, except when the velocity dispersion profile has a negative gradient. In this case, it was not possible to get a good match to both mass and density at $r_1$. 

Our MW disk and bulge potentials increase the SIDM density, shorten the interaction timescale and bring the halo into the phase where the central density is increasing and the core is shrinking. In order to get a better match at $r_1$ than in \citet{oliver18}, we allow for a small negative gradient in the $\sigma_{\rm disp}$ profile using the ansatz $\sigma_{\rm disp,0}(1+\alpha\tanh(r/30\, \rm kpc))$. The value of $30~\rm kpc$ was chosen to be comparable to $r_1$, with the expectation that the required gradients will be small. We decreased $\alpha$ (starting from zero) until the SIDM profile could match on to a NFW profile exactly at $r_1$ (in both mass and density). Thus, the slope of the $\sigma_{\rm disp}$ is automatically set by this procedure. The required value of $\alpha$ is about $-0.2$, which implies $d\sigma_{\rm disp}(r)/dr$ is about $-6\%$ at 10 kpc, consistent with the mild temperature gradient seen in the simulation. For less dense disks, no gradients are required to match the inner and outer profiles at $r_1$ and hence we will revert back to the construction in \citet{manoj16}. 

The inferred analytic profiles for the MW host from the SIDM simulations (DMO and with the embedded potential) are shown in Fig. 5 (dashed lines), and the simulation results are included for reference (solid  lines). The profiles are obtained as described above by matching on to a NFW profile that is fit to the CDM simulation results at $r > 5 \rm kpc$ using the {\tt Colossus} Python package. The agreement is remarkably good for both simulations (with and without the baryonic potential). When the bulge and disk are present, there is excellent agreement between the analytic and simulated profiles in the baryon-dominated region ($r \leq1 \kpc$) and in the asymptotic DM-dominated regime ($r>10 \kpc$); in the transition region between these two regimes the fit is less good but still within about $25\%$.

\bibliography{mybib}

\begin{thebibliography}{61}
\expandafter\ifx\csname natexlab\endcsname\relax\def\natexlab#1{#1}\fi

\bibitem[{{Behroozi}, {Wechsler} \& {Conroy}(2013){Behroozi}, {Wechsler}, \&
  {Conroy}}]{behroozi13}
{Behroozi} P.~S., {Wechsler} R.~H., {Conroy} C., 2013, \apj, 770, 57

\bibitem[{Behroozi, Wechsler \& Wu(2013)Behroozi, Wechsler, \&
  Wu}]{behroozi13a}
Behroozi P.~S., Wechsler R.~H., Wu H.-Y., 2013, The Astrophysical Journal, 762,
  109

\bibitem[{{Behroozi} {et~al}\mbox{.}(2013){Behroozi}, {Wechsler}, {Wu},
  {Busha}, {Klypin}, \& {Primack}}]{behroozi13b}
{Behroozi} P.~S., {Wechsler} R.~H., {Wu} H.-Y., {Busha} M.~T., {Klypin} A.~A.,
  {Primack} J.~R., 2013, \apj, 763, 18

\bibitem[{Bland-Hawthorn \& Gerhard(2016)}]{hawthorn16}
Bland-Hawthorn J., Gerhard O., 2016, Annual Review of Astronomy and
  Astrophysics, 54, 529

\bibitem[{{Boylan-Kolchin}, {Bullock} \& {Kaplinghat}(2011){Boylan-Kolchin},
  {Bullock}, \& {Kaplinghat}}]{boylan11}
{Boylan-Kolchin} M., {Bullock} J.~S., {Kaplinghat} M., 2011, \mnras, 415, L40

\bibitem[{{Boylan-Kolchin}, {Bullock} \& {Kaplinghat}(2012){Boylan-Kolchin},
  {Bullock}, \& {Kaplinghat}}]{boylan12}
---, 2012, \mnras, 422, 1203

\bibitem[{Brooks \& Zolotov(2014)}]{brooks14}
Brooks A.~M., Zolotov A., 2014, The Astrophysical Journal, 786, 87

\bibitem[{Creasey {et~al}\mbox{.}(2017)Creasey, Sameie, Sales, Yu,
  Vogelsberger, \& Zavala}]{creasey17}
Creasey P., Sameie O., Sales L.~V., Yu H.-B., Vogelsberger M., Zavala J., 2017,
  Monthly Notices of the Royal Astronomical Society, 468, 2283

\bibitem[{{Dav{\'e}} {et~al}\mbox{.}(2001){Dav{\'e}}, {Spergel}, {Steinhardt},
  \& {Wandelt}}]{dave2001}
{Dav{\'e}} R., {Spergel} D.~N., {Steinhardt} P.~J., {Wandelt} B.~D., 2001,
  \apj, 547, 574

\bibitem[{Di~Cintio {et~al}\mbox{.}(2017)Di~Cintio, Tremmel, Governato,
  Pontzen, Zavala, Bastidas~Fry, Brooks, \& Vogelsberger}]{dicintio17}
Di~Cintio A., Tremmel M., Governato F., Pontzen A., Zavala J., Bastidas~Fry A.,
  Brooks A., Vogelsberger M., 2017, Monthly Notices of the Royal Astronomical
  Society, 469, 2845

\bibitem[{Elbert {et~al}\mbox{.}(2015)Elbert, Bullock, Garrison-Kimmel, Rocha,
  O\~norbe, \& Peter}]{oliver15}
Elbert O.~D., Bullock J.~S., Garrison-Kimmel S., Rocha M., O\~norbe J., Peter
  A. H.~G., 2015, Monthly Notices of the Royal Astronomical Society, 453, 29

\bibitem[{Elbert {et~al}\mbox{.}(2018)Elbert, Bullock, Kaplinghat,
  Garrison-Kimmel, Graus, \& Rocha}]{oliver18}
Elbert O.~D., Bullock J.~S., Kaplinghat M., Garrison-Kimmel S., Graus A.~S.,
  Rocha M., 2018, The Astrophysical Journal, 853, 109

\bibitem[{Fitts {et~al}\mbox{.}(2017)Fitts, Boylan-Kolchin, Elbert, Bullock,
  Hopkins, O\~{n}orbe, Wetzel, Wheeler, Faucher-Gigu\`{e}re, Keres, Skillman,
  \& Weisz}]{fitts17}
Fitts A. {et~al.}, 2017, Monthly Notices of the Royal Astronomical Society,
  471, 3547

\bibitem[{{Fritz} {et~al}\mbox{.}(2018){Fritz}, {Battaglia}, {Pawlowski},
  {Kallivayalil}, {van der Marel}, {Sohn}, {Brook}, \& {Besla}}]{fritz18}
{Fritz} T.~K., {Battaglia} G., {Pawlowski} M.~S., {Kallivayalil} N., {van der
  Marel} R., {Sohn} S.~T., {Brook} C., {Besla} G., 2018, A\&A, 619, A103

\bibitem[{Fry {et~al}\mbox{.}(2015)Fry, Governato, Pontzen, Quinn, Tremmel,
  Anderson, Menon, Brooks, \& Wadsley}]{fry15}
Fry A.~B. {et~al.}, 2015, Monthly Notices of the Royal Astronomical Society,
  452, 1468

\bibitem[{Garrison-Kimmel {et~al}\mbox{.}(2014)Garrison-Kimmel, Boylan-Kolchin,
  Bullock, \& Kirby}]{kimmel14}
Garrison-Kimmel S., Boylan-Kolchin M., Bullock J.~S., Kirby E.~N., 2014,
  Monthly Notices of the Royal Astronomical Society, 444, 222

\bibitem[{{Garrison-Kimmel} {et~al}\mbox{.}(2018){Garrison-Kimmel}, {Hopkins},
  {Wetzel}, {Bullock}, {Boylan-Kolchin}, {Keres}, {Faucher-Giguere},
  {El-Badry}, {Lamberts}, {Quataert}, \& {Sanderson}}]{kimmel18}
{Garrison-Kimmel} S. {et~al.}, 2018, ArXiv e-prints

\bibitem[{Garrison-Kimmel {et~al}\mbox{.}(2017)Garrison-Kimmel, Wetzel,
  Bullock, Hopkins, Boylan-Kolchin, Faucher-Giguere, Keres, Quataert,
  Sanderson, Graus, \& Kelley}]{kimmel17}
Garrison-Kimmel S. {et~al.}, 2017, Monthly Notices of the Royal Astronomical
  Society, 471, 1709

\bibitem[{Hahn \& Abel(2011)}]{hahn11}
Hahn O., Abel T., 2011, Monthly Notices of the Royal Astronomical Society, 415,
  2101

\bibitem[{{Hernquist}(1990)}]{hernquist90}
{Hernquist} L., 1990, \apj, 356, 359

\bibitem[{Hopkins(2015)}]{hopkins15}
Hopkins P.~F., 2015, Monthly Notices of the Royal Astronomical Society, 450, 53

\bibitem[{Kamada {et~al}\mbox{.}(2017)Kamada, Kaplinghat, Pace, \&
  Yu}]{kamada17}
Kamada A., Kaplinghat M., Pace A.~B., Yu H.-B., 2017, Phys. Rev. Lett., 119,
  111102

\bibitem[{Kaplinghat {et~al}\mbox{.}(2014)Kaplinghat, Keeley, Linden, \&
  Yu}]{manoj14}
Kaplinghat M., Keeley R.~E., Linden T., Yu H.-B., 2014, Phys. Rev. Lett., 113,
  021302

\bibitem[{Kaplinghat, Tulin \& Yu(2016)Kaplinghat, Tulin, \& Yu}]{manoj16}
Kaplinghat M., Tulin S., Yu H.-B., 2016, Phys. Rev. Lett., 116, 041302

\bibitem[{{Katz} \& {White}(1993)}]{katz93}
{Katz} N., {White} S.~D.~M., 1993, \apj, 412, 455

\bibitem[{{Kelley} {et~al}\mbox{.}(2018){Kelley}, {Bullock}, {Garrison-Kimmel},
  {Boylan-Kolchin}, {Pawlowski}, \& {Graus}}]{kelley2018}
{Kelley} T., {Bullock} J.~S., {Garrison-Kimmel} S., {Boylan-Kolchin} M.,
  {Pawlowski} M.~S., {Graus} A.~S., 2018, arXiv e-prints

\bibitem[{Kirby {et~al}\mbox{.}(2014)Kirby, Bullock, Boylan-Kolchin,
  Kaplinghat, \& Cohen}]{kirby14}
Kirby E.~N., Bullock J.~S., Boylan-Kolchin M., Kaplinghat M., Cohen J.~G.,
  2014, Monthly Notices of the Royal Astronomical Society, 439, 1015

\bibitem[{{Koch} {et~al}\mbox{.}(2007){Koch}, {Kleyna}, {Wilkinson}, {Grebel},
  {Gilmore}, {Evans}, {Wyse}, \& {Harbeck}}]{koch07}
{Koch} A., {Kleyna} J.~T., {Wilkinson} M.~I., {Grebel} E.~K., {Gilmore} G.~F.,
  {Evans} N.~W., {Wyse} R.~F.~G., {Harbeck} D.~R., 2007, \aj, 134, 566

\bibitem[{{Mateo}, {Olszewski} \& {Walker}(2008){Mateo}, {Olszewski}, \&
  {Walker}}]{mateo08}
{Mateo} M., {Olszewski} E.~W., {Walker} M., 2008, The Astrophysical Journal,
  675, 201

\bibitem[{McConnachie(2012)}]{mcon12}
McConnachie A.~W., 2012, The Astronomical Journal, 144, 4

\bibitem[{{Miyamoto} \& {Nagai}(1975)}]{miyamoto75}
{Miyamoto} M., {Nagai} R., 1975, \pasj, 27, 533

\bibitem[{{Mu{\~n}oz} {et~al}\mbox{.}(2005){Mu{\~n}oz}, {Frinchaboy},
  {Majewski}, {Kuhn}, {Chou}, {Palma}, {Sohn}, {Patterson}, \&
  {Siegel}}]{munoz05}
{Mu{\~n}oz} R.~R. {et~al.}, 2005, \apjl, 631, L137

\bibitem[{O\~{n}orbe {et~al}\mbox{.}(2014)O\~{n}orbe, Garrison-Kimmel, Maller,
  Bullock, Rocha, \& Hahn}]{onorbe14}
O\~{n}orbe J., Garrison-Kimmel S., Maller A.~H., Bullock J.~S., Rocha M., Hahn
  O., 2014, Monthly Notices of the Royal Astronomical Society, 437, 1894

\bibitem[{Oman {et~al}\mbox{.}(2015)Oman, Navarro, Fattahi, Frenk, Sawala,
  White, Bower, Crain, Furlong, Schaller, Schaye, \& Theuns}]{oman15}
Oman K.~A. {et~al.}, 2015, Monthly Notices of the Royal Astronomical Society,
  452, 3650

\bibitem[{{Papastergis} {et~al}\mbox{.}(2015){Papastergis}, {Giovanelli},
  {Haynes}, \& {Shankar}}]{papa15}
{Papastergis} E., {Giovanelli} R., {Haynes} M.~P., {Shankar} F., 2015, A\&A,
  574, A113

\bibitem[{{Papastergis} \& {Shankar}(2016)}]{papa16}
{Papastergis} E., {Shankar} F., 2016, A\&A, 591, A58

\bibitem[{Peter {et~al}\mbox{.}(2013)Peter, Rocha, Bullock, \&
  Kaplinghat}]{peter13}
Peter A. H.~G., Rocha M., Bullock J.~S., Kaplinghat M., 2013, Monthly Notices
  of the Royal Astronomical Society, 430, 105

\bibitem[{{Planck Collaboration}(2015)}]{planck15}
{Planck Collaboration}, 2015, AA, 594, A13

\bibitem[{{Read}, {Walker} \& {Steger}(2018){Read}, {Walker}, \&
  {Steger}}]{read2018}
{Read} J.~I., {Walker} M.~G., {Steger} P., 2018, \mnras, 481, 860

\bibitem[{{Ren} {et~al}\mbox{.}(2018){Ren}, {Kwa}, {Kaplinghat}, \&
  {Yu}}]{ren18}
{Ren} T., {Kwa} A., {Kaplinghat} M., {Yu} H.-B., 2018, ArXiv e-prints

\bibitem[{{Robertson}, {Massey} \& {Eke}(2017){Robertson}, {Massey}, \&
  {Eke}}]{robertson17}
{Robertson} A., {Massey} R., {Eke} V., 2017, \mnras, 467, 4719

\bibitem[{{Robertson} {et~al}\mbox{.}(2018){Robertson}, {Massey}, {Eke},
  {Tulin}, {Yu}, {Bah{\'e}}, {Barnes}, {Bower}, {Crain}, {Dalla Vecchia},
  {Kay}, {Schaller}, \& {Schaye}}]{robertson2018}
{Robertson} A. {et~al.}, 2018, \mnras, 476, L20

\bibitem[{Robles {et~al}\mbox{.}(2017)Robles, Bullock, Elbert, Fitts,
  Gonz?lez-Samaniego, Boylan-Kolchin, Hopkins, Faucher-Gigu?re, Kere?, \&
  Hayward}]{robles17}
Robles V.~H. {et~al.}, 2017, Monthly Notices of the Royal Astronomical Society,
  472, 2945

\bibitem[{Rocha {et~al}\mbox{.}(2013)Rocha, Peter, Bullock, Kaplinghat,
  Garrison-Kimmel, O{\~n}orbe, \& Moustakas}]{rocha13}
Rocha M., Peter A. H.~G., Bullock J.~S., Kaplinghat M., Garrison-Kimmel S.,
  O{\~n}orbe J., Moustakas L.~A., 2013, Monthly Notices of the Royal
  Astronomical Society, 430, 81

\bibitem[{Sameie {et~al}\mbox{.}(2018)Sameie, Creasey, Yu, Sales, Vogelsberger,
  \& Zavala}]{omid18}
Sameie O., Creasey P., Yu H.-B., Sales L.~V., Vogelsberger M., Zavala J., 2018,
  Monthly Notices of the Royal Astronomical Society, 479, 359

\bibitem[{Sawala {et~al}\mbox{.}(2016)Sawala, Frenk, Fattahi, Navarro, Bower,
  Crain, Vecchia, Furlong, Helly, Jenkins, Oman, Schaller, Schaye, Theuns,
  Trayford, \& White}]{sawala16}
Sawala T. {et~al.}, 2016, Monthly Notices of the Royal Astronomical Society,
  457, 1931

\bibitem[{{Simon} \& {Geha}(2007)}]{simon07}
{Simon} J.~D., {Geha} M., 2007, The Astrophysical Journal, 670, 313

\bibitem[{{Smith} {et~al}\mbox{.}(2015){Smith}, {Flynn}, {Candlish},
  {Fellhauer}, \& {Gibson}}]{smith15}
{Smith} R., {Flynn} C., {Candlish} G.~N., {Fellhauer} M., {Gibson} B.~K., 2015,
  \mnras, 448, 2934

\bibitem[{Spergel \& Steinhardt(2000)}]{spe00}
Spergel D.~N., Steinhardt P.~J., 2000, Phys. Rev. Lett., 84, 3760

\bibitem[{{Springel}, {Frenk} \& M.(2006){Springel}, {Frenk}, \&
  M.}]{springel16}
{Springel} V., {Frenk} C.~S., M. W. S.~D., 2006, \nat, 440, 1137

\bibitem[{Tollerud, Boylan-Kolchin \& Bullock(2014)Tollerud, Boylan-Kolchin, \&
  Bullock}]{tollerud14}
Tollerud E.~J., Boylan-Kolchin M., Bullock J.~S., 2014, Monthly Notices of the
  Royal Astronomical Society, 440, 3511

\bibitem[{Trujillo-Gomez {et~al}\mbox{.}(2011)Trujillo-Gomez, Klypin, Primack,
  \& Romanowsky}]{trujillo11}
Trujillo-Gomez S., Klypin A., Primack J., Romanowsky A.~J., 2011, The
  Astrophysical Journal, 742, 16

\bibitem[{Tulin \& Yu(2018)}]{tulin17}
Tulin S., Yu H.-B., 2018, Physics Reports, 730, 1 , dark matter
  self-interactions and small scale structure

\bibitem[{Vogelsberger {et~al}\mbox{.}(2014{\natexlab{a}})Vogelsberger, Genel,
  Springel, Torrey, Sijacki, Xu, Snyder, Nelson, \& Hernquist}]{vogel14a}
Vogelsberger M. {et~al.}, 2014{\natexlab{a}}, Monthly Notices of the Royal
  Astronomical Society, 444, 1518

\bibitem[{{Vogelsberger}, {Zavala} \& {Loeb}(2012){Vogelsberger}, {Zavala}, \&
  {Loeb}}]{vogel12}
{Vogelsberger} M., {Zavala} J., {Loeb} A., 2012, \mnras, 423, 3740

\bibitem[{Vogelsberger {et~al}\mbox{.}(2014{\natexlab{b}})Vogelsberger, Zavala,
  Simpson, \& Jenkins}]{vogel14b}
Vogelsberger M., Zavala J., Simpson C., Jenkins A., 2014{\natexlab{b}}, Monthly
  Notices of the Royal Astronomical Society, 444, 3684

\bibitem[{{Walker} \& {Pe{\~n}arrubia}(2011)}]{walker09}
{Walker} M.~G., {Pe{\~n}arrubia} J., 2011, \apj, 742, 20

\bibitem[{Wetzel {et~al}\mbox{.}(2016)Wetzel, Hopkins, {Ji-hoon},
  Faucher-Giguere, Keres, \& Quataert}]{wetzel16}
Wetzel A.~R., Hopkins P.~F., {Ji-hoon} K., Faucher-Giguere C.-A., Keres D.,
  Quataert E., 2016, The Astrophysical Journal Letters, 827, L23

\bibitem[{{Wolf} {et~al}\mbox{.}(2010){Wolf}, {Martinez}, {Bullock},
  {Kaplinghat}, {Geha}, {Mu{\~n}oz}, {Simon}, \& {Avedo}}]{wolf10}
{Wolf} J., {Martinez} G.~D., {Bullock} J.~S., {Kaplinghat} M., {Geha} M.,
  {Mu{\~n}oz} R.~R., {Simon} J.~D., {Avedo} F.~F., 2010, \mnras, 406, 1220

\bibitem[{{Zavala}, {Vogelsberger} \& {Walker}(2013){Zavala}, {Vogelsberger},
  \& {Walker}}]{zavala13}
{Zavala} J., {Vogelsberger} M., {Walker} M.~G., 2013, \mnras, 431, L20

\bibitem[{Zolotov {et~al}\mbox{.}(2012)Zolotov, Brooks, Willman, Governato,
  Pontzen, Christensen, Dekel, Quinn, Shen, \& Wadsley}]{zolotov12}
Zolotov A. {et~al.}, 2012, The Astrophysical Journal, 761, 71

\end{thebibliography}

\bsp
\label{lastpage}
\end{document}